# *Statdepth*: a package for analysis of functional and pointcloud data using statistical depth

**Julian Lehrer**
UC Santa Cruz Genomics Institute

**Mircea Teodorescu**
UC Santa Cruz Genomics Institute

**Yohei Rosen**
UC Santa Cruz Genomics Institute

**Abstract**

Laboratory scientists are well equipped with statistical tools for univariate data, yet many phenomena of scientific interest are time-variant or otherwise multidimensional. Functional data analysis is one way of approaching such data: by representing these more complex data as single data points in a mathematical space of functions. The mathematical concept of functional depth provides a notion of centrality which allows for descriptive statistics and some comparative statistics on these data. Here, we present **statdepth**, a Python package for functional depth-based analyses which naturally extends familiar single-data-point L-statistics and related methods to time-variant data trajectories or multidimensional data.

## 1. Introduction: On Functional Data Analysis

In recent years, especially in the field of statistics for biology, numerous scientific instruments collect functional data (López-Pintado and Romo (2009)).

For example, in the field of biology, a single replicate of gene expression data derived from RNA sequencing might entail tens of thousands of scalar dimensions. Biological parameters such as neuroelectrical activity, hormone levels, biological tissue composition or morphology change dynamically and in nonlinear manners poorly approximated by pointwise comparisons.

Naturally, we are interested in analysis of this data, and in particular, distinguishing between the curves when they arise from replicate experiments. Although the notion of statistical depth was first introduced (Tukey 1975) for multivariate points in order to generalize the idea of order statistics such as the median, functional depth extends this notion to functional data (López-Pintado and Romo 2009). From functional depth arises the idea of band depth, a functional depth statistic based on the geometry of the functional curves. This provides an outward ordering of our set of curves based on centrality. Additionally, we define the most central curve to be the median. Building on this new definition of depth for functional data, (Flores, Lillo, and Romo 2018) provide homogeneity testing measures for functional samples.

There are currently no packages for functional depth calculations in Python (Van Rossum and Drake 2009). There are two packages in R. Among other functional statistics methods, **fda.usc** implements functional depth calculations in R, including band depth (limited to $J = 2$), but does not implement homogeneity testing or visualization methods. (Febrero-Bande and de la Fuente (2012)) **ddalpha** provides extensive depth measures for pointcloud data, but lacks any depth measures for functional data. (Pokotylo, Mozharovskyi, and Dyckerhoff (2019)) In the **statdepth** package, we provide methods for depth calculations of both functional and pointcloud data using a variety of depth measures. Additionally, we implement homogeneity testings methods and depth visualizations.

Our package **statdepth** is built using the fundamental statistical libraries **pandas** and **numpy** for statistics, and **plot.ly** for visualization. **Pandas** is a widely used package for data analysis in `Python`, particularly in bioinformatics. It is fast, scalable, and contains a well-documented API. For this reason, **statdepth** was built to interface with **Pandas**, and its methods return **Pandas** objects whenever possible. There are four core classes in **statdepth**. The `PointcloudDepth` and `FunctionalDepth` classes, which handle computations for pointcloud and functional depth, and the `PointcloudHomogeneity` and `FunctionalHomogeneity` classes, which handle calculating homogeneity statistics between pairs of point cloud and functional data sets.

# 2. Mathematical Background

## 2.1. The Band Depth

As a motivating example, first consider a univariate statistical distribution over the real line. Consider independent random variables $x_1, \ldots, x_n$ sampled from this distribution. Using the standard linear ordering of the real numbers, we can calculate such L-statistics Daniell (1920) as the sample median or the interquartile range. These give estimators of a central value and a dispersion parameter respectively which are both easily calculable and interpretable by a statistical layperson. Assuming a unimodal distribution, the median is furthermore an easily interpretable estimator of the maximum density of the distribution and might be thought of as the "most representative" value of the distribution.

Now consider that instead of univariate points, we have independent random variables drawn from a statistical distribution over a space of functions $f : D \to R$, where $D, R \subseteq \mathbb{R}$. Informally, let us say that $D$ is a time domain such that the graphs of our functions are discrete or continuous-time time variant curves. Let curves $X_1, \ldots, X_n$ be realizations of these independent random variables. Informally, we might say that $X_1, \ldots, X_n$ are time series scientific observations generated by some underlying process. We seek to define a statistic for our $X$'s such that we might define similar L-statistics for these time-series data.

Informally, define the $j$-band depth of a curve $X$ is the proportion possible of combinations of $j$ curves for which the envelope that contain $X$. Intuitively, if $n$ is sufficiently large and this proportion is low, then the curve $X$ must be divergent in either shape or magnitude and therefore not well representative of the underlying experimental group. If this proportion is large, then the curve is central in the set and is representative of the underlying distribution.

Note that this is not a linear ordering of the curves $X$ in magnitude: first, it is instead a centrality ordering and second it is generally a partial ordering.

Nevertheless, López-Pintado and Romo (2009) show that this centrality-based partial ordering also allows for L-statistics to be calculated. And indeed, inspection of the definitions shows that in the degenerate case of real-valued single point data (which may be represented by functions $f : 1 \to \mathbb{R}$) then the functional depth centrality ordering arrives at the exact same median, interquartile range and other L-statistics that are generated by the magnitude based ordering conventionally used.

This definition of the median is intuitive and theoretically justified, satisfying the standard properties for a depth measure: affine invariance, maximality at center, monotonicity relative to the deepest point, and vanishing at infinity López-Pintado and Romo (2009), Serfling and Zuo (2000).

The band depth is also particularly useful for identifying subtle outliers, since the shape of the curve is also relevant when considering its containment in the delimiting bands. Therefore, a curve can be an outlier without having any values that are particularly large.

We begin by defining the "band" in the real plane $\mathbb{R}^2$ defined by a set of two or more curves.

Consider the domain of $x_1, ..., x_n$ to be the compact subset $I \subset \mathbb{R}$ of the real line. Consider a set $x_1, ..., x_n$ of real-valued curves $x_i : I \to \mathbb{R}$. Let $G(x)$ be the graph in $\mathbb{R}^2$ of $x$. The band defined by the curves $x_{i_1}, ...x_{i_k}$, where $i_1, .., i_k$ is some subsequence of indices $1, ..., n$, is the region in the plane defined by

$$V(x_{i_1}, ..., x_{i_k}) = \left\{(t, y) \in \mathbb{R}^2 \mid t \in I, \min_{r=1,...,k} x_{i_r}(t) \leq y \leq \max_{r=1,...,k} x_{i_r}(t)\right\}$$

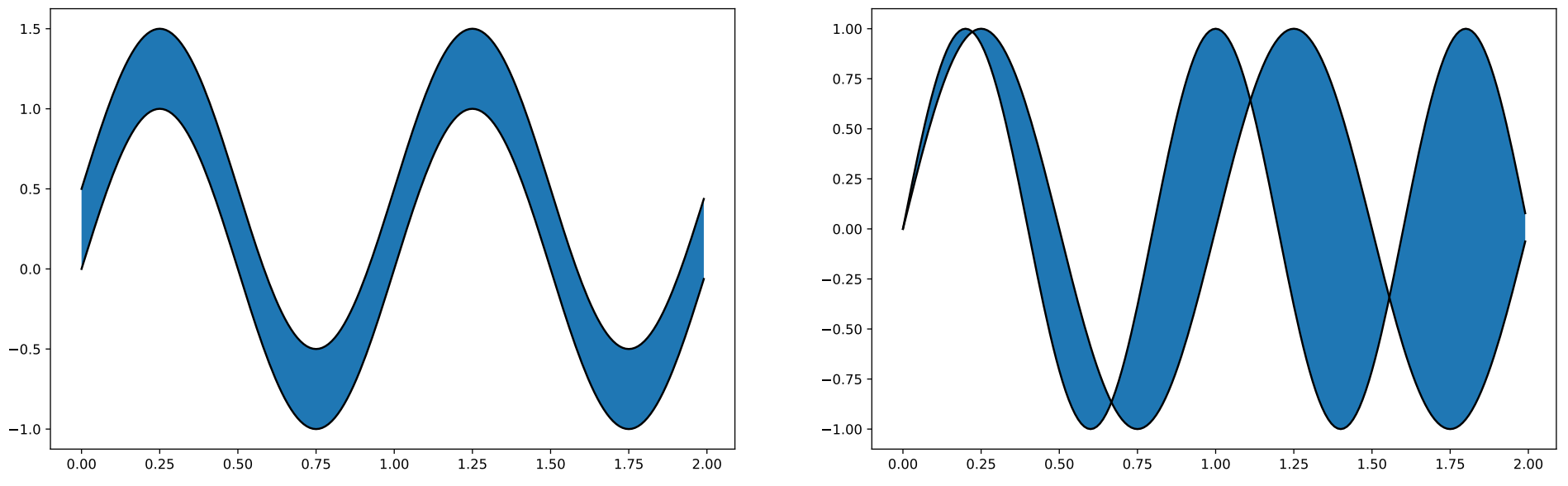

(a) Bands defined in $\mathbb{R}^2$

Figure 1: Example of bands with two similar curves (left) and two dissimilar curves (right)

Then for any $x$ in our sample $x_1, ..., x_n$, the proportion of bands determined by $j \geq 2$ different curves $x_{i_1}, ..., x_{i_j}$ that contain the curve $x$ is

$$S_n^j(x) = \binom{n}{j}^{-1} \sum_{1 \leq i_1 \leq ... \leq i_j \leq n} \mathbb{I}\left\{G(x) \subset V(x_{i_1}, ..., x_{i_j})\right\}$$

Where we are summing over all possible length $j$ subsequences of $x_1, ..., x_n$.

We define the **band depth** of a function $x$ given the sample $x_1, ..., x_n$ to be

$$D(x \mid x_1, ..., x_n) \equiv S_{n,J}(x) = \sum_{j=2}^{J} S_n^j(x)$$

Where $J \geq 2$.

Having a notion of centrality allows us to compute, for example, a central range such as an interquartile range or a median. The latter is defined as the curve with the largest depth value,

$$\text{Median}(x_1, ..., x_n) \equiv x_{deepest} = \arg\max_{x \in x_1, ... x_n} S_{n,J}(x)$$

and the interquartile range is defined by the half of the set of curves with the highest depth.

## 2.2. Modified Band Depth

Suppose our $x_1, ..., x_n$ are quite irregular (or irregular at some small interval in the domain $I$) in that that each $x_i$ crosses some $x_j$ at at least one point.

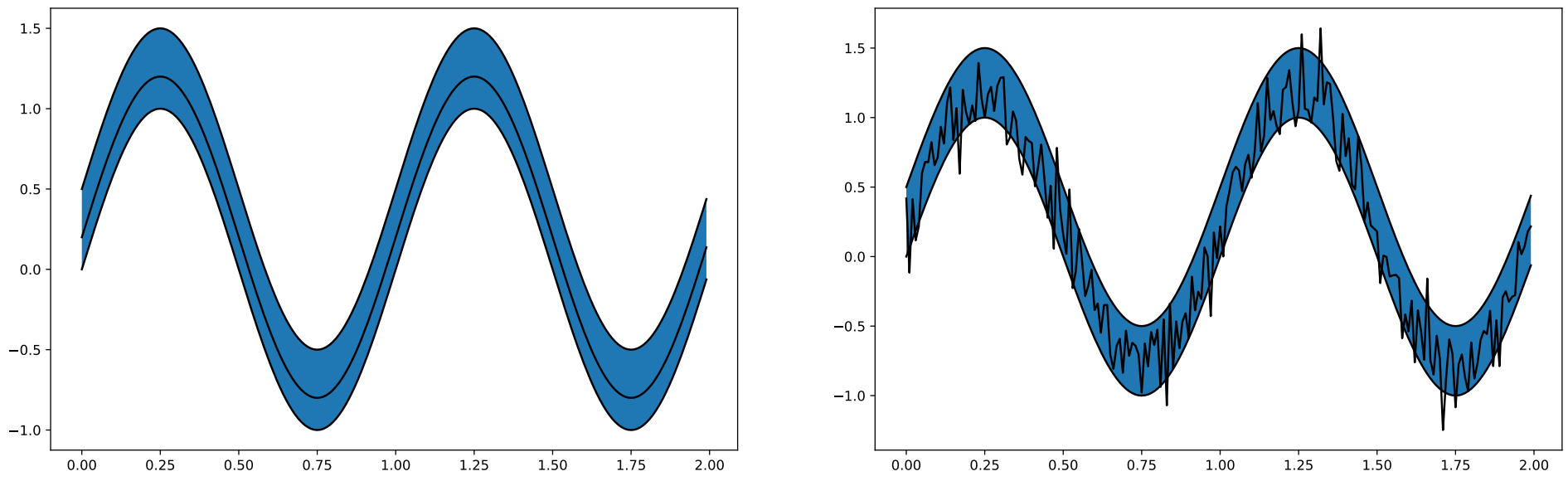

(a) Functional curve in a band defined in $\mathbb{R}^2$

Figure 2: Example of a regular curve contained by a band (left) and an irregular curve escaping containment at small intervals (right)

Then, since our definition of containment $\mathbb{I}\left\{G(x) \subset V(x_{i_1}, ..., x_{i_j})\right\}$ only considers if the graph of $x$ is *entirely* contained within the band formed by $x_{i_1}, ..., x_{i_j}$, a set of with many dissimilar curves will have a lot of zero depth values, and the generated ordering may not be particularly useful. This may occur in the case of data obtained from noisy instruments, or an inhomogeneous source of data.

Therefore, we define a **modified band depth** López-Pintado and Romo (2009). For $j \geq 2$ define

$$A(x \mid x_{i_1}, ..., x_{i_j}) = \left\{t \in I \mid \min_{r=i_1,...,i_j} x_r(t) \leq x(t) \leq \max_{r=i_1,...,i_j} x_r(t)\right\}$$

Which are the set of points in $I$ where the function $x$ is in the band determined by $x_{i_1}, ..., x_{i_j}$. Let $\lambda$ be the Lebesgue measure in $I$, and therefore

$$\lambda_r(A(x \mid x_{i_1}, ..., x_{i_j})) := \frac{\lambda(A(x \mid x_{i_1}, ..., x_{i_j}))}{\lambda(I)}$$

gives the "proportion of time" that $x$ is in the band formed by $x_{i_1}, ..., x_{i_j}$. This allows a natural generalization of $S_n^j(x)$, given by

$$GS_n^j(x) = \sum_{1 \leq i_1 \leq ... \leq i_j \leq n} \lambda_r(A(x \mid x_{i_1}, ..., x_{i_j}))$$

Note that if $x$ is always inside the band, the value of $\lambda_r(A(x \mid x_{i_1}, ..., x_{i_j}))$ is one which returns us to the original value using the indicator function.

We now define the modified band depth for a curve $x$ in $x_1, ..., x_n$ as

$$MBD_{n,J}(x) \equiv GS_{n,J}(x) = \sum_{j=2}^{J} GS_n^i(x), J \geq 2$$

The main difference between the band depth and the modified band depth is that the first is more sensitive to the shape of the curve, while the latter depends more on its magnitude. This is because if a ill-shaped curve sits close to the center of the functional sample, it will still have low depth value since small intervals will not be contained in many bands López-Pintado and Romo (2009).

## 2.3. Simplicial Depth

Band depth provides an intuitive definition for the most central curve of a set of a data, but only works well where each function is univariate. Consider the case of multivariate curves wherein we have a set $x_1, ..., x_n$ of multivariate curves $x_i : I \longrightarrow \mathbb{R}^d$. Immediately, our definition of containment given above breaks down, because we need to define containment in an interval where the interval is high-dimensional.

For this reason, *simplex depth* was developed by (López-Pintado, Jornsten *et al.* 2007). First, consider the case where the time domain is a single point $x \in \mathbb{R}^d$ with respect to the sample $X = \{x_1, ..., x_n\}$ where $x_i \in \mathbb{R}^d$. We partition the set into $\binom{n}{d+1}$ simplices and consider the proportion of simplices that contain $x$.

We define the simplicial depth of a point $x \in \mathbb{R}^d$ with respect to the sample $x_1, ..., x_n$ to be

$$SD(x \mid x_1, ..., x_n) = \binom{n}{d+1}^{-1} \sum_{1 \leq i_1 \leq ... \leq i_{d+1} \leq n} \mathbb{I}\left\{x \subset Simplex(x_{i_1}, ..., x_{i_{d+1}})\right\}$$

To generalize both simplicial depth and band depth to multivariate functional data, consider $x_i : D \longrightarrow \mathbb{R}^d$ where $D$ is the domain. For example, we've collected two features of an experiment every thirty minutes for a day.

Then, we define a simplicial band depth of $x$ with respect to $x_1, ..., x_n$ as

$$sBD(x \mid x_1, ..., x_n) = \binom{n}{d+1}^{-1} \sum_{1 \leq i_1 \leq ... \leq i_{d+1} \leq n} \mathbb{I}\left\{\left(\sum_{t \in D} \mathbb{I}\left\{x(t) \subset Simplex(x_{i_1}(t), ..., x_{i_{d+1}}(t))\right\}\right) = |D|\right\}$$

That is, for each subsequence, we calculate the indicator function of $x(t)$ being contained in all $d + 1$ dimensional simplex formed by $x_{i_1}(t), ..., x_{i_{d+1}}(t)$ for each $t \in D$, and zero otherwise. Naturally, we also extend the modified band depth with the modified simplicial band depth given by

$$sMBD(x \mid x_1, ..., x_n) = \binom{n}{d+1}^{-1} \sum_{1 \le i_1 \le ... \le i_{d+1} \le n} \frac{1}{|D|} \left( \sum_{t \in D} \mathbb{I}\left\{ x(t) \subset Simplex(x_{i_1}(t), ..., x_{i_{d+1}}(t)) \right\} \right)$$

For each subsequence, we calculate the proportion of $d+1$ dimensional simplices that contains $x(t)$, $t \in D$, instead of checking that it is contained in *all* $d+1$ dimensional simplices.

## 2.4. Homogeneity Testing

Statistical depth also allows for homogeneity between two functional samples $F$ and $G$. That is, we would like to determine if these two samples have been generated by the same process; equivalently, if they have the same sample distribution.

Consider an arbitrary sample $g$ and a set of data $F$, where $g$ is not necessarily a member of $F$. Let $d_F(g)$ to be the depth of $g$ with respect to $F \cup \{g\}$.

Suppose $g \in G$, and define

$$\mathfrak{D}_F(\mathfrak{G}) = \arg\max_{g \in G} d_F(g)$$

That is, $\mathfrak{D}_F(\mathfrak{G})$ is the sample in $g$ which maximizes the value $d_F(g)$. Alternatively, we can consider this to be the deepest function of $G$ with respect to $F$.

(Flores *et al.* 2018) defines three homogeneity coefficients whose values reach their extrema if our sets $F$ and $G$ are generated by the same generating process.

So, define

$$\mathbf{P_1}(F, G) = d_F\left(\mathfrak{D}_G G\right)$$

Where $\mathfrak{D}_G(G)$ is the element of $G$ with the highest depth value. Therefore, if we compute how deep this estimator with respect to $F$, we can say that the greater this depth, the less likely the two samples come from different generating processes (Flores *et al.* 2018).

Similarly, (Flores *et al.* 2018) defines

$$\mathbf{P_2}(F, G) = |\mathbf{P_1}(F, G) - \mathbf{P_1}(F, F)|$$

Observe that the $\mathbf{P_2}$ statistic is a normalization of $\mathbf{P_1}$. If the nature of the generating processes which produces sample $F$ produces shallow depths, then $P_1(F, F)$ would be an estimate of the maximum of these depths. Therefore, the sets $F$ and $G$ are likely produced by the same process as the $P_2$ coefficient gets close to zero. (Flores *et al.* 2018) provides a more detailed review of functional homogeneity coefficients.

## 2.5. Computation of Band Depth

The computational cost of calculating the band depth depends on the sample size $n$ and the parameter $j = 2, ..., J$. Since the band depth for a curve $x$ requires iteration over all possible subsequences of lengths $j = 2, ..., J$, of which there are $\binom{n}{2} + ...\binom{n}{J}$. The time complexity of the exact computation is $\mathcal{O}(\binom{n}{J})$. This makes band depth computationally intractable for large $n$ (Sun, Genton, and Nychka 2012).

To solve this problem, (López-Pintado and Romo 2009) propose the following resampling algorithm to provide a computationally efficient approximation.

Given $K > 1$,

1. Randomly assign $n$ curves into $K$ blocks $B_1, ..., B_k$ of size $\sim n/k$

2. Compute depth values (using any valid functional depth measure) for each $x$ with respect to $B_i$, denote this $D(x \mid B_i)$

3. The sampled depth value of $x$ is $D(x) = K^{-1} \sum_{i=1}^{K} D(x \mid B_i)$

Therefore, our computational cost for computing the depth of a curve $x$ using resampling is $K\binom{n/K}{2}$, which is smaller than $\binom{n}{2}$ for large $n$ and $K > 1$, supposing that $J = 2$. This method is approximate, so there is a trade-off between time saved and accuracy. However, as the size of our data increases, we are free to select a larger $K$ since the number of samples in each block increases proportionally to the size of our data, so the estimate in each block is closer to the true depth value (Sun *et al.* 2012).

# 3. Using the statdepth package

## 3.1. Core methodologies

Methods for functional and for point cloud data are encapsulated in two classes, `FunctionalDepth` and `PointcloudDepth`, which operate very similarly. Internally, they store both the original data and its depth. We implement useful methods for evaluating depth, plotting the data, and reporting statistics as listed in the table below. Across both classes, the `relax` upon initialization corresponds to standard (`relax=False`) or modified (`relax=True`) depth. The `K` parameter sets the block size for the resampling method. This should be used for large data, since exact computation becomes intractable with a large number of curves 2.5 (Sun *et al.* 2012). A table illustrating the core arguments and methods for both the `FunctionalDepth` and `Pointcloud Depth` is below. For `FunctionalDepth`, the `data` input is a list of `DataFrames` since in the case of multivariate time series data, each functional curve is it's own `DataFrame`

| Parameter | Purpose | Data Type |
|---|---|---|
| `data` | Functional curves to calculate depth for | List of `DataFrames` |
| `J` | $J$ parameter for band depth calculation, controls the number of bands we sum to 2. | Integer $\geq 2$ |
| `K` | $K$ parameter to decide block size in the case of the resampling method. | Integer $\geq 1$. By default this is `None` and computes exact depth |
| `containment` | Containment definition for band depth. The default is `r2`, which denotes the standard band containment for functional curves. For multivariate curves, this uses simplicial depth across each time point. See 2. | One of `simplex`, `r2` for functional data, and one of `simplex`, `oja`, `mahalanobis`, `l1` for pointcloud. |
| `relax` | Functional depth parameter where `True` calculates modified band depth and `False` calculates standard band depth. | Boolean, default is `False` |
| `quiet` | Boolean parameter indicating whether to display calculation progress during depth calculations | Boolean, default is `False` |
| `deep_check` | Boolean parameter indicating whether to perform data validation before depth calculations. | Boolean, default is `False` |

Table 1: Arguments for both `FunctionalDepth` and `PointcloudDepth` classes.

| Method | Purpose |
|---|---|
| `ordered` | Sort and return the curves by their functional depth |
| `deepest(n=1)` | Return the depths for the $n$ deepest curves. For $n = 1$, this defines the median. |
| `outlying(n=1)` | Return depths for the $n$ most outlying curves. In the case of ties, both curves are returned. |
| `drop_outlying_data(n=1)` | Return the original set of functional curves with the $n$ most outlying trimmed |
| `get_deepest_data(n=1)` | Return the $n$ deepest curves |
| `plot_deepest(n=1)` | Plot all the functional curves, with the $n$ deepest marked in red. |
| `plot_outlying(n=1)` | Plot all the functional curves with the $n$ outlying marked in red |

Table 2: Methods for both `FunctionalDepth` and `PointcloudDepth` classes. Default arguments are in parenthesis under the method column

For homogeneity testing, we define the classes `FunctionalHomogeneity` and `PointcloudHomogeneity`. These methods are for calculating the coefficients $P_i(F, G), i = 1, ..4$.

| Parameter | Purpose | Data Type |
|---|---|---|
| `F` | Functional curves that define the set $F$ | List of `DataFrames` |
| `G` | Functional curves that define the set $F$ | List of `DataFrames` |
| `method` | Determines homogeneity coefficient to calculate | One of `p1`, `p2`, `p3`, `p4` |
| `J` | $J$ parameter for band depth calculation, controls the number of bands we sum to 2. | Integer $\geq 2$ |
| `K` | $K$ parameter to decide block size in the case of the resampling method. | Integer $\geq 1$. By default this is `None` and computes exact depth |
| `containment` | Containment definition for band depth. The default is `r2`, which denotes the standard band containment for functional curves. For multivariate curves, this uses simplicial depth across each time point. See 2. | One of `simplex`, `r2` for functional data, and one of `simplex`, `oja`, `mahalanobis`, `l1` for pointcloud. |
| `relax` | Functional depth parameter where `True` calculates modified band depth and `False` calculates standard band depth. | boolean |
| `quiet` | Boolean parameter indicating whether to display calculation progress during depth calculations | Boolean, default is `False` |
| `deep_check` | Boolean parameter indicating whether to perform data validation before depth calculations. | Boolean, default is `False` |

Table 3: Arguments for both `FunctionalHomogeneity` and `PointcloudHomogeneity` classes.

### 3.2. statdepth for univariate functional data

Here, we demonstrate code examples for **statdepth**, and how a complete functional analysis pipeline may look. **statdepth** assumes each column of an $m \times n$ `DataFrame` is be a functional curve sampled at $m$ time points, for a total of $n$ curves. Below, we show a simple example of reading in functional data and visualizing the two deepest curves.

```
>>> df
     f_0  f_1  f_2  f_3  f_4  f_5
x_0    1    2    3  6.0    9    8
x_1    2    4    4  7.0    9    8
x_2    3    5    4  6.5   12   10
x_3    2    6    2  6.0   11   10
x_4    1    2    1  7.0   11    9
>>> type(df)
```

```
<class 'pandas.core.frame.DataFrame'>
```

Then to get the depth of each curve using band depth (by default), we run

```
>>> from statdepth import FunctionalDepth

>>> d = FunctionalDepth([df], J=2, relax=False).ordered()
>>> d
f_3     0.400000
f_5     0.266667
f_2     0.200000
f_1     0.200000
f_4     0.000000
f_0     0.000000
dtype: float64
```

To visualize the $n$ deepest and most outlying curves, we run

```
>>> d.plot_deepest(N)
>>> d.plot_outlying(N)
```

whose results are seen in the following figure. To demonstrate a minimal example with **statdepth**, we generate a seed curve $f_0$ by sampling 20 values from the uniform distribution over $(0, 1]$, then generate perturbed data such that each $f_i = x_i f_0$ where $x_i \sim Unif(0, 1)$.

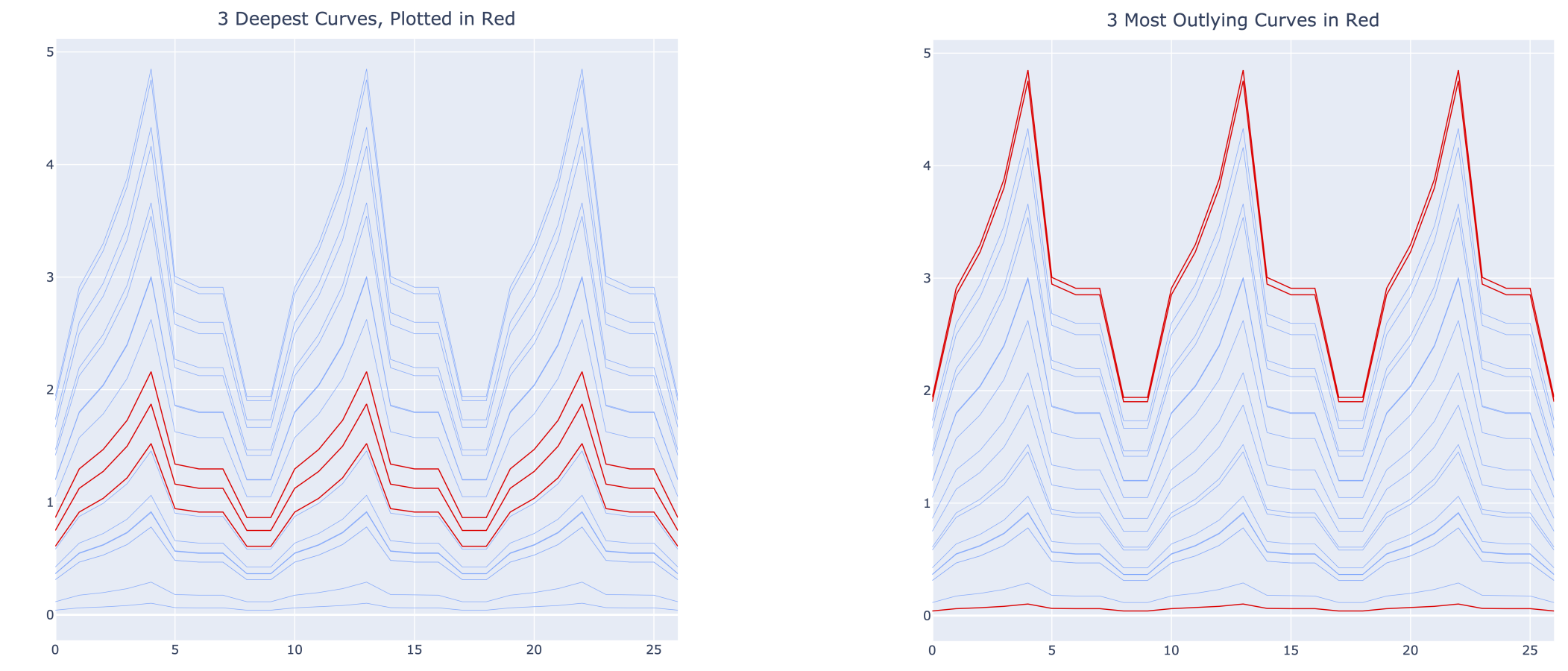

(a) 3 Deepest Curves

(b) 3 Outlying Curves

Figure 3: Example functional data, 3 deepest and most outlying curves marked in red.

### 3.3. statdepth for multivariate functional data

In the multivariate functional case, a single `DataFrame` is not able to define multiple multivariate data in an intuitive way. Therefore, each $m \times p$ multivariate functional curve is $m$ samples in $p$ dimensions. That is, each column is a dimension and each row is a point from $\mathbb{R}^p$.

So suppose our multivariate curves are given by the following `DataFrame`s

```
    >>> df1
      size  co_amount  weight
00:00     0          2       2
00:30     1          0       3
01:00     1          3       2
01:30     3          0       3
02:00     3          3       1
02:30     1          1       0

>>> df2
      size  co_amount  weight
00:00     3          2       3
00:30     0          3       2
01:00     1          3       2
01:30     2          2       0
02:00     0          1       2
02:30     1          3       3

>>> df3
...
```

Then we order the curves using (simplicial depth by default) by running

```
>>> from statdepth import FunctionalDepth

>>> FunctionalDepth([df1, df2, df3, ... , df6], containment='simplex', J=2, relax=True)
2    0.333333
1    0.333333
5    0.166667
0    0.166667
4    0.000000
3    0.000000
dtype: float64
```

### 3.4. statdepth for pointcloud data

In the pointcloud case, our $m \times p$ DataFrame is $m$ points in $\mathbb{R}^p$. Suppose our data is the following along two dimensions

```
>>> df
          0         1
0  0.873179  0.828111
```

```
1  0.368512  0.024619
2  0.927522  0.348593
3  0.481917  0.748796
4  0.980515  0.954392
```

Similarly, in the three dimensional case we have an additional column. Then the following gives the ordering of the curves, using simplicial depth.

```
>>> from statdepth import PointcloudDepth
>>> PointcloudDepth(df, containment='simplex')
0    0.703605
1    0.239076
2    0.458779
3    0.456768
4    0.258959
dtype: float64
```

The same visualization methods apply to the `PointcloudDepth` object for points in $\mathbb{R}^2$ and $\mathbb{R}^3$, whose results can be seen in the following figure.

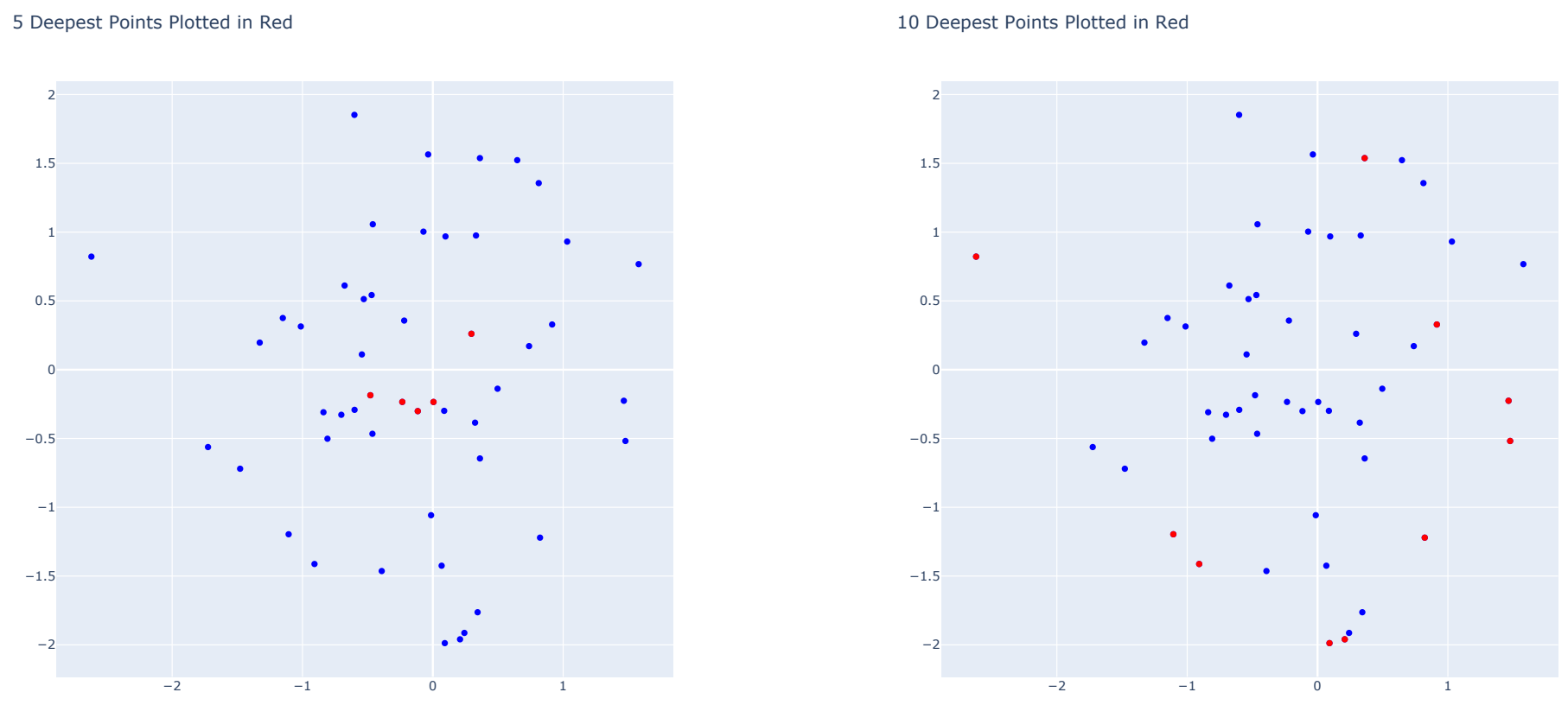

(a) 3 Deepest Curves (b) 3 Outlying Curves

Figure 4: Example of deepest and outlying points sampled from a uniform distribution over 0, 1, with the 3 deepest points are marked in red (left), and the 10 most outlying points marked in red (right).

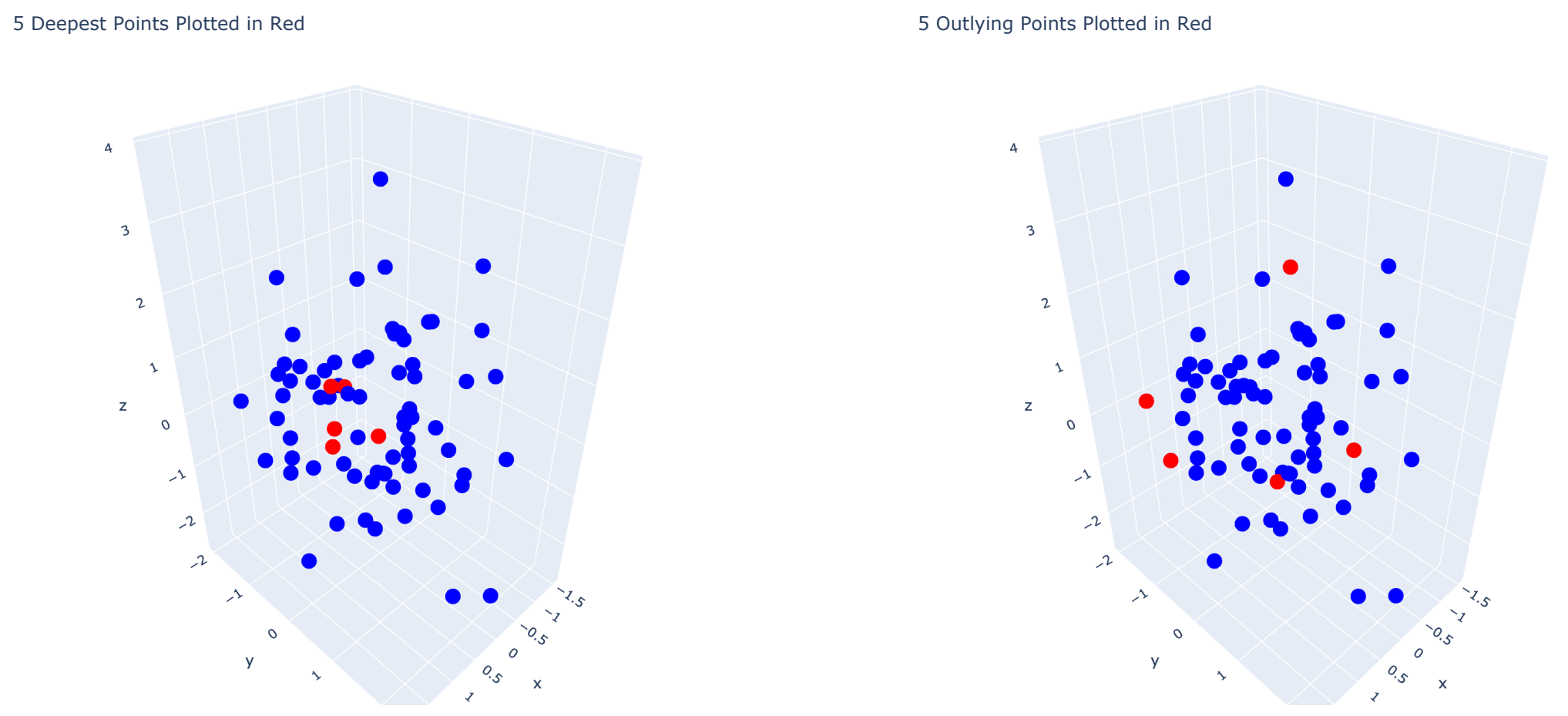

(a) 5 Deepest Points

(b) 5 Outlying Points

Figure 5: Example of deepest and outlying points from 50 points from a standard normal distribution, with the 3 deepest points are marked in red (left), and the 10 most outlying points marked in red (right).

Additionally, a complete ordering of the points based on color can be viewed with the `plot_depths` method for points in $\mathbb{R}^2$ and $\mathbb{R}^3$, as seen below.

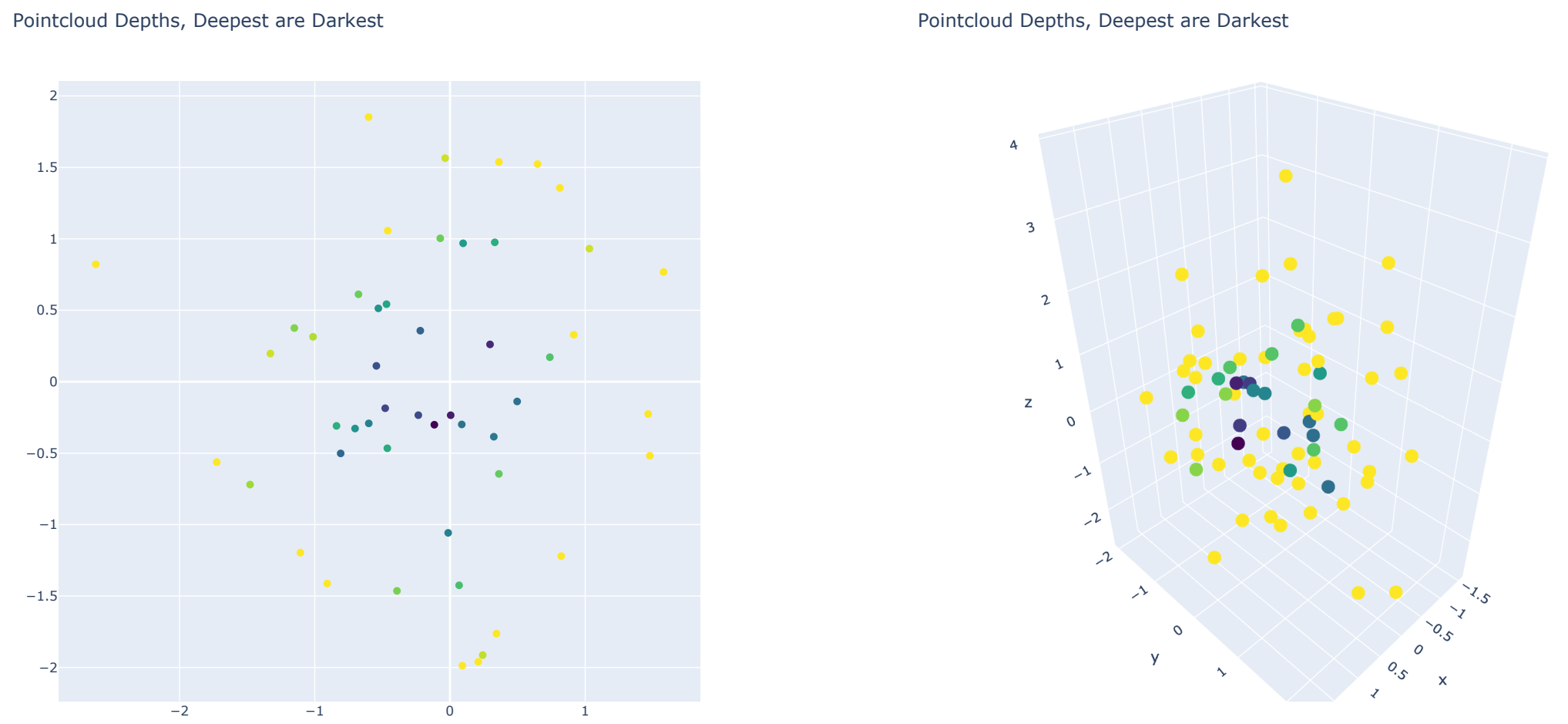

(a) Complete ordering for 75 standard normal sampled points in $\mathbb{R}^2$

(b) Complete ordering for 75 standard normal sampled points in $\mathbb{R}^3$

Figure 6: Example of deepest and outlying points sampled from a uniform distribution over $0, 1$, with the 3 deepest points are marked in red (left), and the 10 most outlying points marked in red (right).

# 4. A Complete Pipeline with statdepth

We consider data from five rat feeding trials conducted as part of the GRACE project (Huang and Walker 2019). This experiment tested how rat growth changed with different feedings. There are five experimental groups. Group 1 is the control, group 2 has 11% GMO food, group 3 has 33% GMO food, group 4 is conventional food 1, and group 5 is conventional food 2.

This experiment was performed across four trials, but for brevity we consider two. Here, we demonstrate calculating the $P2$ (see 2.4) homogeneity coefficient (lower is better) between the sets of curves, and visualizing the $n = 3$ deepest.

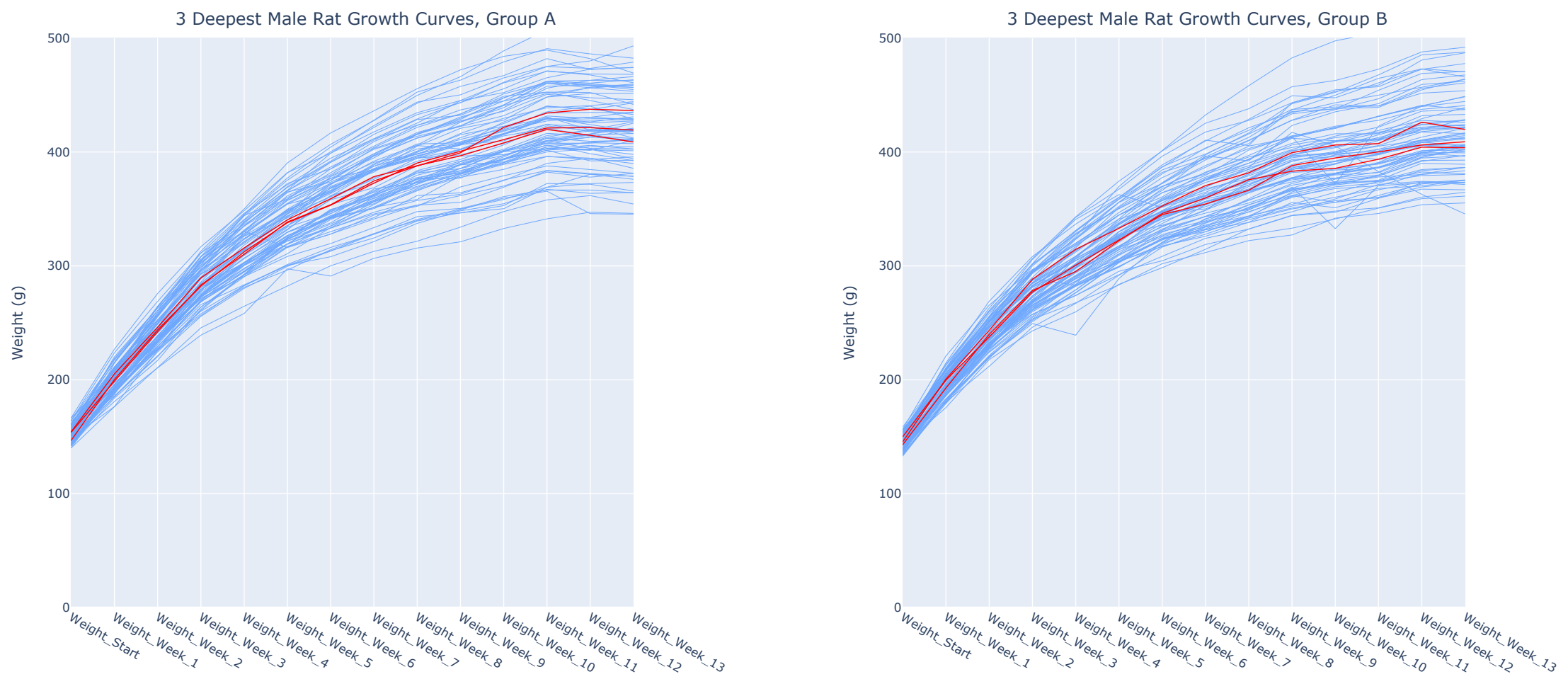

(a) Curves from trial $A$ (b) Curves from trial $B$

Figure 7: $n = 3$ deepest and outlying curves, colored in red, from trials $A$ and $B$ of the GRACE rat feeding project Huang and Walker (2019)

Then to calculate the $P2$ homogeneity coefficient where `df_a`, `df_b` contain the growth curves from trials $A$ and $B$, respectively, we use

```
from statdepth.homogeneity import FunctionalHomogeneity
>>> FunctionalHomogeneity([df_a], [df_b], K=10, J=2, relax=True, method='p2')
0.04
```

We can now safely assume that the functional samples are distributionally similar (Flores *et al.* 2018), because the median curves are similar in both groups. We can now perform further analysis on the combined data from trials $A$ and $B$. Of course, a natural analysis would be to compare experimental groups. First, we visualize the $n = 3$ deepest curves for each experimental group.

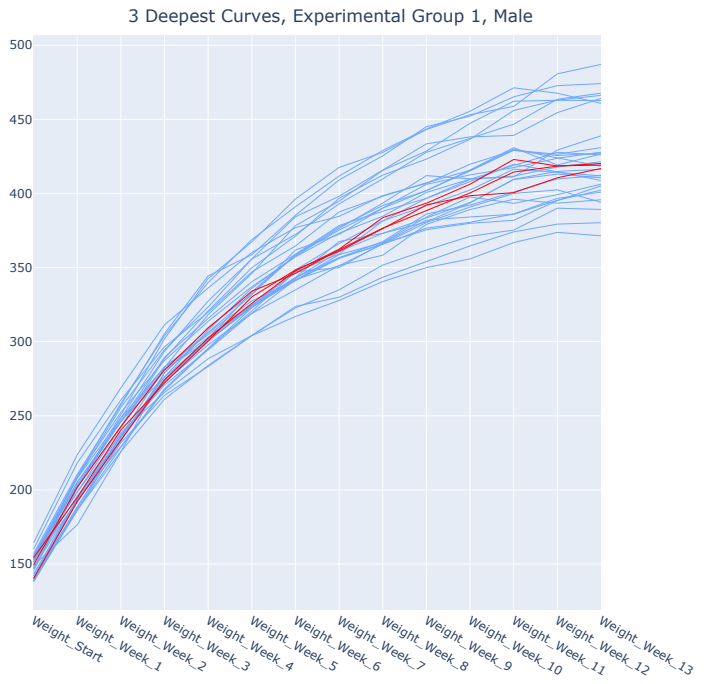

(a) Experimental group 1

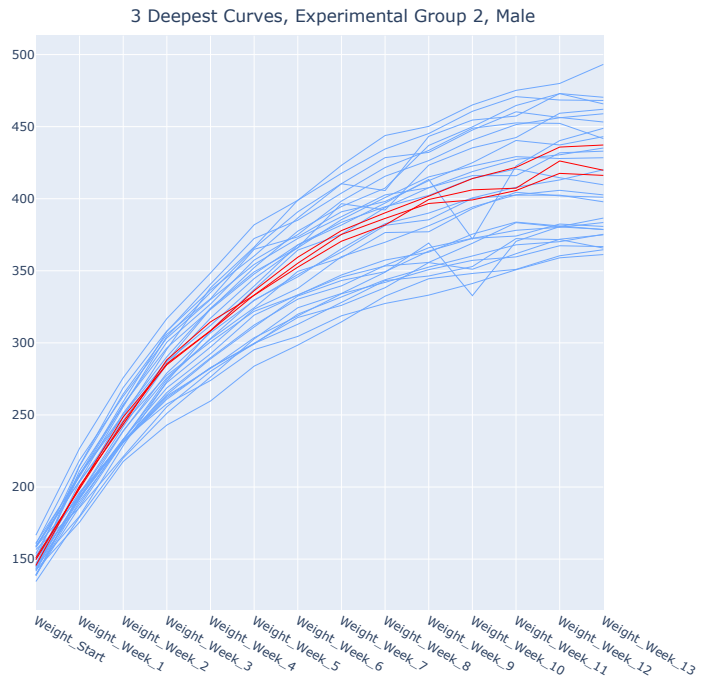

(b) Experimental group 2

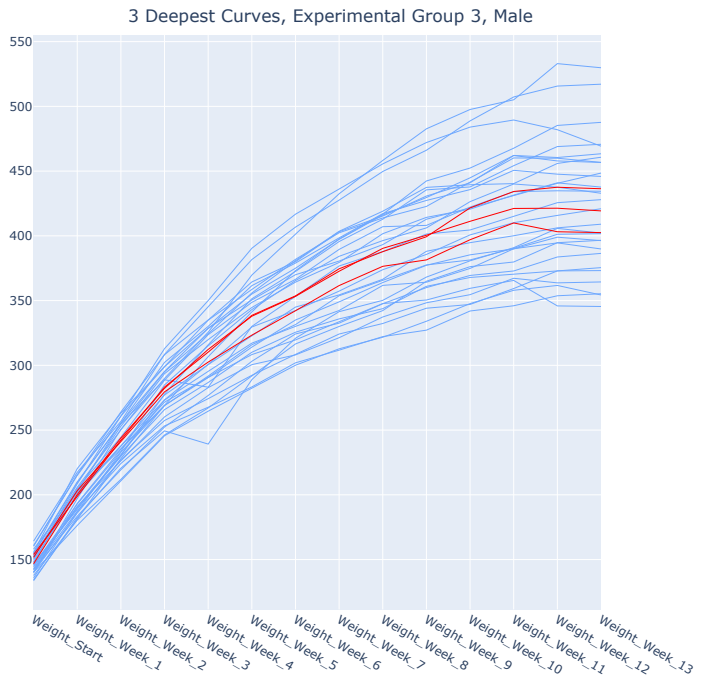

(c) Experimental group 3

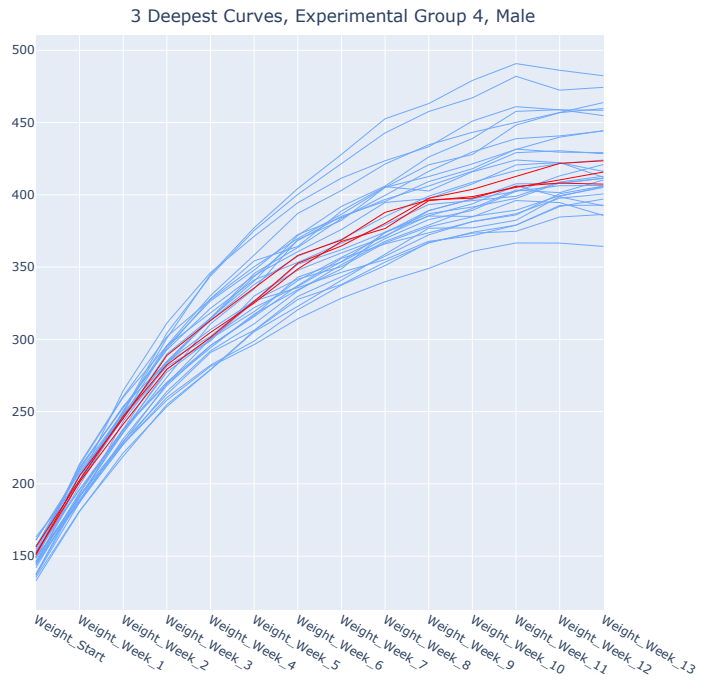

(d) Experimental group 4

Figure 8: 3 Deepest Curves colored in red for each experimental group $i = 1, 2, 3, 4$

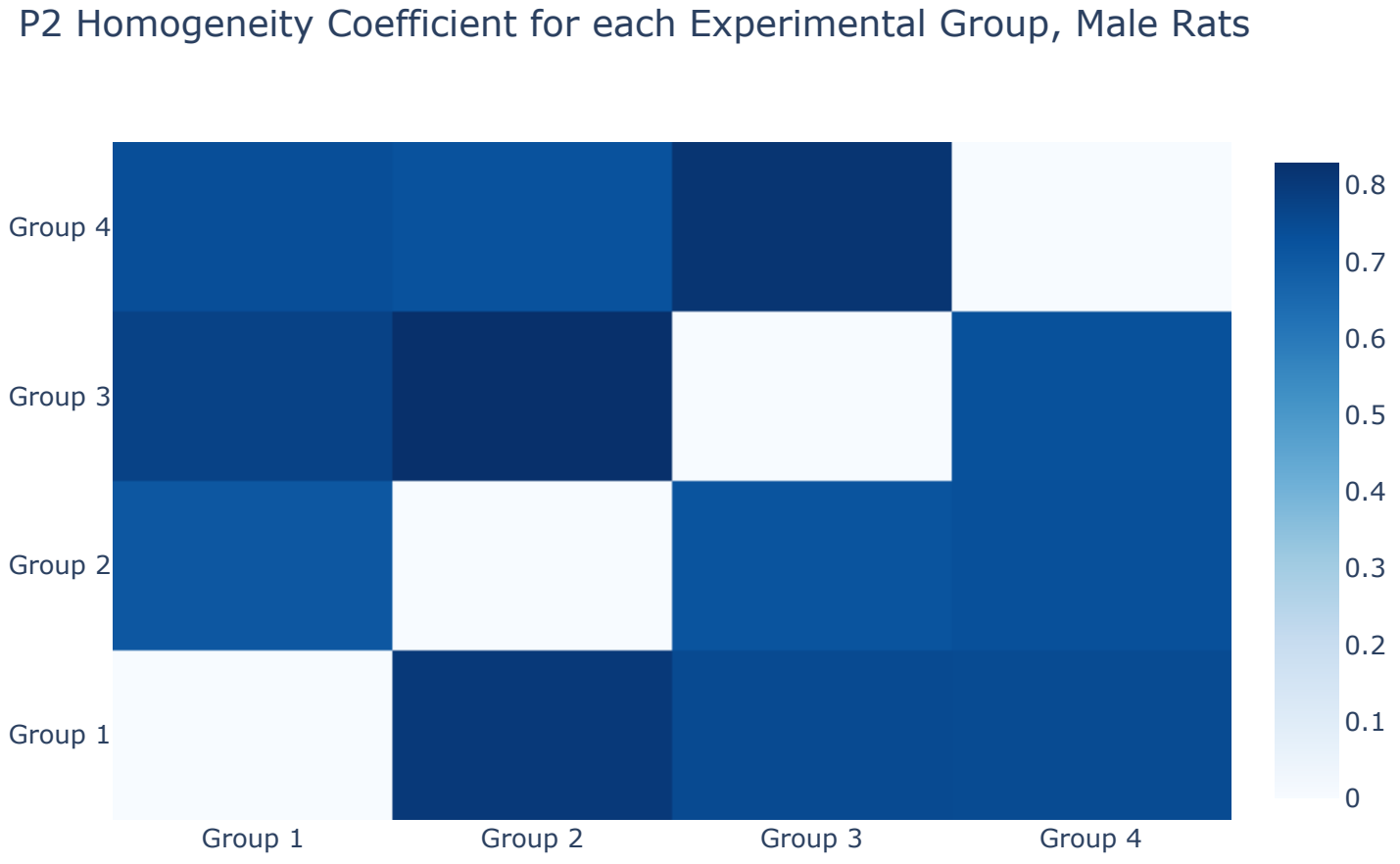

Figure 9: P2 Homogeneity Coefficient between each experimental group

Finally, we can calculate the homogeneity coefficients between each group, and visualize this as a heatmap. Note that $P_i(F, G)$ and $P_i(G, F)$ are not necessarily expected to be equal (Flores *et al.* 2018).

## 5. Conclusion

In this paper, we have introduced the package **statdepth** and demonstrated it's usefulness in the analysis time series or other multivariate data arising from scientific experiments. The concept of band depth which we rely on is a particularly useful statistic, yet no package for its computation in Python had been previously developed. We implement band depth and simplicial depth calculations for functional data, pointcloud data, and homogeneity testing between functional groups. Additionally, visualization tools are provided for each.

As we continue to develop **statdepth**, we plan to implement parallelization methods for inexact depth calculated by the resampling methods described in 2.5. We encourage any contributions and/or bug reports to our codebase, which can be found https://github.com/braingeneers/statdepth. Finally, we welcome any questions or suggestions through email to the main package developer, Julian Lehrer.

## Acknowledgments

We thank Dr. Alex Pang for his help with an early version of this work. This work was supported by the Schmidt Futures Foundation award number SF 857. Additionally, we thank David Parks for his astute knowledge of Python package development, and his guidance in developing good software.